\documentclass[12pt,a4paper,oneside]{article}
\usepackage[left=2.5cm,right=2.5cm,top=2.2cm,bottom=2.3cm]{geometry}
\usepackage[utf8]{inputenc}
\usepackage[english]{babel}
\usepackage{amsmath}
\usepackage{amsfonts}
\usepackage{amssymb}
\usepackage{makeidx}
\usepackage[colorlinks]{hyperref}
\usepackage{color}
\usepackage{graphicx}
\usepackage{graphics}
\usepackage{lmodern}
\usepackage[toc]{appendix}
\usepackage{authblk}

\newcommand{\email}[1]{\normalsize\href{mailto:#1}{#1} }
\newcommand{\nb}{\nonumber}

\author{Abdelkader YANALLAH\footnote{Private email. \email{yanallahabdelkader@hotmail.com}}}
\title{On 2d gravity and parallel transport \\ field equation on quantum sphere.}
\date{}
\affil{\footnotesize Laboratoire de Physique Th\'eorique et de l'Interaction Rayonnement-Mati\`ere (LPTHIRM),
\\ Universit\'e BLIDA1, BLIDA, Algeria.\\ \email{yanallahabdelkader@univ-blida.dz}}

\begin{document}
\maketitle

\begin{abstract}
In this work we have obtained the exact quantum expressions for the compenents of the Levi Cevita connection, the Ricci tensor and the scalar curvature, generalizing  those of \cite{CTZX08} for a spherical surface via the noncommutative Moyal star product, and we have established equations describing quantum effect on the geodesic flow equation or auto parallel fields equations. These later are solved for the zero  and the first orders of the quantum parameter $\alpha$ when $y$-symmetry is assumed. We expressed the general system of equations in terms  of Fourier modes indexed by the integer  $p=1,2,...,\infty$ to understand the interdependence of $y$ modes oscillations.
\end{abstract}


\section{Introduction}
Since the work of N. Seiberg and E. Witten \cite{SW} there has been a great interest on describing  spacetime in the framework of noncommutative geometry in the aim to understand its effect on fundamental theories of physics.
One of the main purposes  of noncommutativity of the spacetime is to resolve the gravitational singularities problem at very high energy scales when quantum effects are considered. Profound   theoretical and experimental investigations were developed to link noncommutative parameters to experimental facts \cite{SD}, \cite{SDF} and \cite{KLV}. Seiberg-Witten map provides a bridge between the ring of functions space, with quantum parameters and classical product, and  the noncommutative space provided by associative star product called Moyal product. This approach was fruitful  to quantize  gravity theories  considered as gauge theories, see for instance \cite{CTZX08} where an interesting formalism was constructed to deal with noncommutative Riemannian surfaces  or reference \cite{Koba} where a $\Theta$-Twisted gravity was basically constructed on  $\Theta$-twisted general coordinate transformations and local Lorentz invariance. In the context of gauge group $ISO(3,1)$ we cite the work of A.H Chamseddine \cite{Cham} where a deformed gravity is constructed using the Seiberg-Witten map. The generalization of noncommutative gauge theories to the case of orthogonal
and symplectic groups are discussed in \cite{Djabari}. Noncommutative QFT with relativistic invariance were studied using twisted Poincaré symmetry in \cite{chaichian}. We can not cover all topics  of  gravity on noncommutative spaces and quantum gravity but we just bring intention to reference \cite{wess} which is nice alternative to \cite{CTZX08}, and to \cite{Modesto} to show how a noncommutative quantum spacetime with minimal length scale behave as $2d$ manifold. And for general aspects of noncommutative gravity we cite the following collection: \cite{Nicolini}, \cite{Moffat}, \cite{Calmet2}, \cite{Calmet3},  \cite{Garay},\cite{Hinchliffe}, \cite{Freidel} and \cite{Garcia}.

On the other side, a  deformed Schwarzschild solution in noncommutative gauge is computed and red quantum shift effect was estimated in \cite{chaichian2}. Authors of \cite{Calmet} think that spacetime is lower-dimensional at very short distances and hence quantum black holes produced at the LHC or in cosmic rays scattering live in lower dimensions.
Speaking about black holes lead us to two things among others, the no hair conjecture introduced by Wheeler \cite{wheeler} related to extractable physical information (mass, spin ,charge...) from black holes  and quantum spheres since the horizon of static or spining  black holes are almost spherical surfaces experiencing very strong gravity and should then be treated in the framework of quantum gravity \cite{Nicolini}. Spheres have also  relation  to an algebraic topology theorem, known as hairy theorem \cite{Renteln}, which makes restriction on the profile of existing fields  on the sphere. Studying physics in the proximity of black hole horizon asks for intervening quantum spacetime or equivalently quantum gravity effects.

In this work we have made some specific choices to  treat the impact of quantum effects on geodesic trajectories around the black hole horizon assumed as a perfect $2d$-sphere. But since the notion of trajectory in the realm of  quantum physics is a very fuzzy concept we substitute trajectories on the sphere considered as noncommutative  surface by the flow field associated  to geodesic lines submitted to deformed $2d$-gravity via the quantum parameter $h$. The geometric  quantization concerns only the angular coordinates ($x=\theta ,\; y=\phi$) which are the latitude and the longitude of the sphere. The radial part transverse to horizon is decoupled from the problem of quantization. We have principally used the Moyal star product, based on Heinsenberg canonical quantization, and the mathematical formalism constructed in \cite{CTZX08} to achieve exact quantum construction of the $2d$-gravity proprieties and as well to define the quantum version of auto parallel transport field equation.

The organization of this paper is as follows: In section \ref{25} we recall the definition of the quantum  star product defined on $2d$ quantum space as well as some useful proprieties. We choose a convenient basis of functions   for the study of the $2d$ sphere geometry. In section \ref{97}  we introduce the noncommutative $2d$-sphere definition and establish its quantum deformed metric and hence derive the exact expressions of  quantum version  of the  Ricci tensor $R_{ij}$ and the scalar curvature $R$. In section  \ref{210} we define first the noncommutative parallel transport equation by replacing the standard commutative product of functions  by the Moyal star product in the commutative version of geodesic equation. Then, 
 the study is specified to auto parallel case  where the flow equation is resolved under the $y$-symmetry assumption for the zero order and the first order of quantum perturbation parameter $\alpha=\tanh (h)$. The generalization of the flow system equations to arbitrary fields $V^\mu (x,y)$ is then  given  
where  Fourier series analysis in the $y$ direction is exploited to compute  the $p$ 
modes of quantum flow equation.   We conclude this work 
 with few comments and perspectives. In the appendix 
 we collect further expressions on star product and Fourier modes calculation used along this work. Several notations  are used in this paper, thus we have set: 
 $c_n=\cos (n y)$, $s_n=\sin (n y)$, ${\rm ch}_m=\cosh (n h)$, ${\rm sh}_m=\sinh (n h)$ and $\partial_1=\frac{\partial }{\partial x},\;\partial_2=\frac{\partial }{\partial y}$.  The rotation invariance  about the north-south axis designes what we call $y$-symmetry in section \ref{210}.
\section{The Weyl-Moyal product on $2d$ sphere}
\label{25}
Let $X\equiv(x^\mu)\equiv(x,y)$ and $W\equiv(w^\mu)\equiv(u,v)$ represent the coordinates of two points on the two dimensional sphere $S$. $f(X)$ and $g(X)$ are two generic smooth functions on $S$.
The Weyl-Moyal product, denoted by the symbol $\star$, is defined by the following definition
\begin{equation}\label{48}
(f\star g)(X)=\lim_{W \rightarrow X} \exp({\cal V})f(X)g(W),
\end{equation}
with the quadratic derivative operator given by
\begin{equation}
  {\cal V} =  \frac{i}{2}\theta^{\mu\nu}\frac{\partial^2}{\partial x^\mu\partial w^\nu},\qquad\qquad \mu,\nu=1,2.
\end{equation}
where $\theta^{\mu\nu} = -\theta^{\nu\mu} $ are the deformation parameters of the moyal product. We will use $\frac{i}{2}\theta^{12}=\textit{h}$ as short notation. In two dimensions ${\cal V}$ reduces to
\begin{equation}
{\cal V}=\textit{h}\left[\frac{\partial^2}{\partial x \partial v}-\frac{\partial^2}{\partial y \partial u}\right].
\end{equation}
This star product, indexed by the parameter $h$, have the following proprieties:
\begin{eqnarray}
  f\star_h g &=& g\star_{-h}f, \label{41}\\
  f(x)\star_h [g_1(x)g_2(y)] &=& g_1(x)[f(x)\star_h g_2(y)], \\
  (f_1(x) f_2(y))\star_h g(y) &=& (f_1(x)\star_h g(y))f_2(y),
\end{eqnarray}
which can be proved term by term after  expansion of (\ref{48}) according to the power of  $h$. 
Let introduce the set of the basic functions $f_m$, suitable for the geometry of sphere surface, by the following expressions
\begin{equation}\label{68}
   \begin{array}{lcr}
     f_1 = \sin(x) \sin(y), &\quad & f_2 =\sin(x) \cos(y), \\
     f_3 = \cos(x) \sin(y), & & f_4 = \cos(x) \cos(y).
   \end{array}
\end{equation}
We note here that $f_1 f_4=f_2 f_3$ and for any re-scaling $\hat\sigma X:=(\sigma x,\widetilde{\sigma}y)$ of the coordinate $X=(x,y)$ we have
\begin{equation}
f(\hat\sigma X)\star_h g(\hat\sigma X)=(f\star_{h'}g)(X),
\end{equation}
where $h'=\sigma \widetilde{\sigma}h$. The set (\ref{68}) possesses, under the shift transformation $(x,y)\rightarrow (\frac{\pi}{2}-x,\frac{\pi}{2}-y)$, the permutation propriety
\begin{equation}\label{58}
\left\lbrace f_1\longleftrightarrow f_4,\qquad f_2\longleftrightarrow f_3 \right\rbrace.
\end{equation} 
The expanded operator:
\begin{equation}
\exp ({\cal V})= \sum^{\infty}_{n=0}\sum^{n}_{p=0}\frac{h^n(-1)^p}{p!(n-p)!}\frac{\partial^{2n}}{\partial x^{n-p}\partial v^{n-p}\partial y^p\partial u^p}
\end{equation}
is left invariant under the same transformation shift \footnote{$\exp ({\cal V})$ is also invariant under the transformation
\begin{equation}
\{h\longrightarrow -h,\quad x\longleftrightarrow y,\quad u\longleftrightarrow v\}
\end{equation}
which acts on $f_m $ as follow 
\begin{equation}
f_1\longleftrightarrow f_1,\quad f_4\longleftrightarrow f_4,\quad f_2\longleftrightarrow f_3\end{equation}
}. This remark together with (\ref{58}) will reduce star product computation.
Writing a recursive  Mathematica code for $\exp ({\cal V})$ we can prove that the Moyal products $f_n\star f_m$ using (\ref{48}) reduce, for $m=1,\cdots 4$ and $m_0=5-m$, to
\begin{eqnarray}\label{67}
f_m\star f_m &=& {f_m}^2 \cosh^2(h)- {f_{m_0}}^2 \sinh ^2(h),\\
f_1\star f_2\; &=&\left({f_1} \cosh (h)-{f_4} \sinh (h)\right) \left({f_2} \cosh (h)-{f_3} \sinh (h)\right),\\
f_1\star f_3 \;&=&\left({f_1} \cosh (h)+{f_4} \sinh (h)\right) \left({f_3} \cosh (h)+{f_2} \sinh (h)\right),\\
f_1\star f_4 \;&=&{f_1} {f_4}+\left({f_2}^2-{f_3}^2\right) \sinh (h) \cosh (h).\label{71}
\end{eqnarray}
The remaining 9 star products are obtained using prescriptions (\ref{41}) and (\ref{58}). 
On the other hand for any smooth one variable functions  $F(x)$ and $G(y)$ we can obtain by formal computation
\begin{equation}\label{Ffm}
F(x)\star f_m=\frac{1}{2}F(x+ih)(f_m+i\,p \,f_{m_2})+\frac{1}{2}F(x-ih)(f_m-i\,p\,f_{m_2}),
\end{equation}
\begin{equation}\label{fmF}
f_m\star G(y) =\frac{1}{2}G(y+ih)(f_m+i\, s \,f_{m_3})+\frac{1}{2}G(y-ih)(f_m-i\,s\,f_{m_3}),
\end{equation}
where $p=(-1)^m$, $m_2=m-p$, $m_3=5-m_2$ and $s=(m-m_3)/2$. Notice  if $h$ is a real parameter and $F$ and  $G$ are real-valued functions  then (\ref{Ffm}) and (\ref{fmF}) are also reals functions.
\section{The noncommutative sphere}\label{97}
First let ${\cal A}$ be a noncommutative $h$-deformation of the algebra of smooth functions on a local chart of the sphere $S^2$. To define the noncommutative surface  of sphere  we introduce the vector $\Lambda_h$ as an element of the space ${\cal A}^3$ provided with generalized noncommutative scalar product $[a,b,c]\cdot[a',b',c']=a\star a'+b\star b'+c\star c'$ with euclidean signature, where $a,b,c,a',b'$ and $c'$ are functions  belonging to ${\cal A}$. The expression of $\Lambda_h$ is 
\begin{equation}\label{103b}
\Lambda_h=\textrm{A}\,f_2\,e_1+\textrm{A}\,f_1\,e_2+\textrm{B}\,\cos(x)\,e_3,
\end{equation}
with $\{e_1,e_2,e_3\}$ is the canonical basis of ${\cal A}^3$. The coefficients $\textrm A$ and $\textrm B$ are determined by the constraint defining the quantum surface i.e. $\Lambda_h\cdot\Lambda_h=1$. Using  (\ref{67})-(\ref{71}) it is easy to evaluate them and the result is \cite{CTZX08}
\begin{equation}\label{103}
    \begin{array}{cc}
       \textrm{A} = {\cosh(h)}^{-1}, \quad & \quad \textrm{B} =  \textrm{A}  {\sqrt{\cosh(2h)}}.
     \end{array}
\end{equation}
Since $\textrm{A} $ is different from $\textrm{B} $ when $h\neq 0$, the non commutative sphere is viewed as classical ellipsoid with a symmetry rotation along the $z$-axis.
   
\begin{figure}[t]
\begin{center}
\includegraphics[scale=0.8]{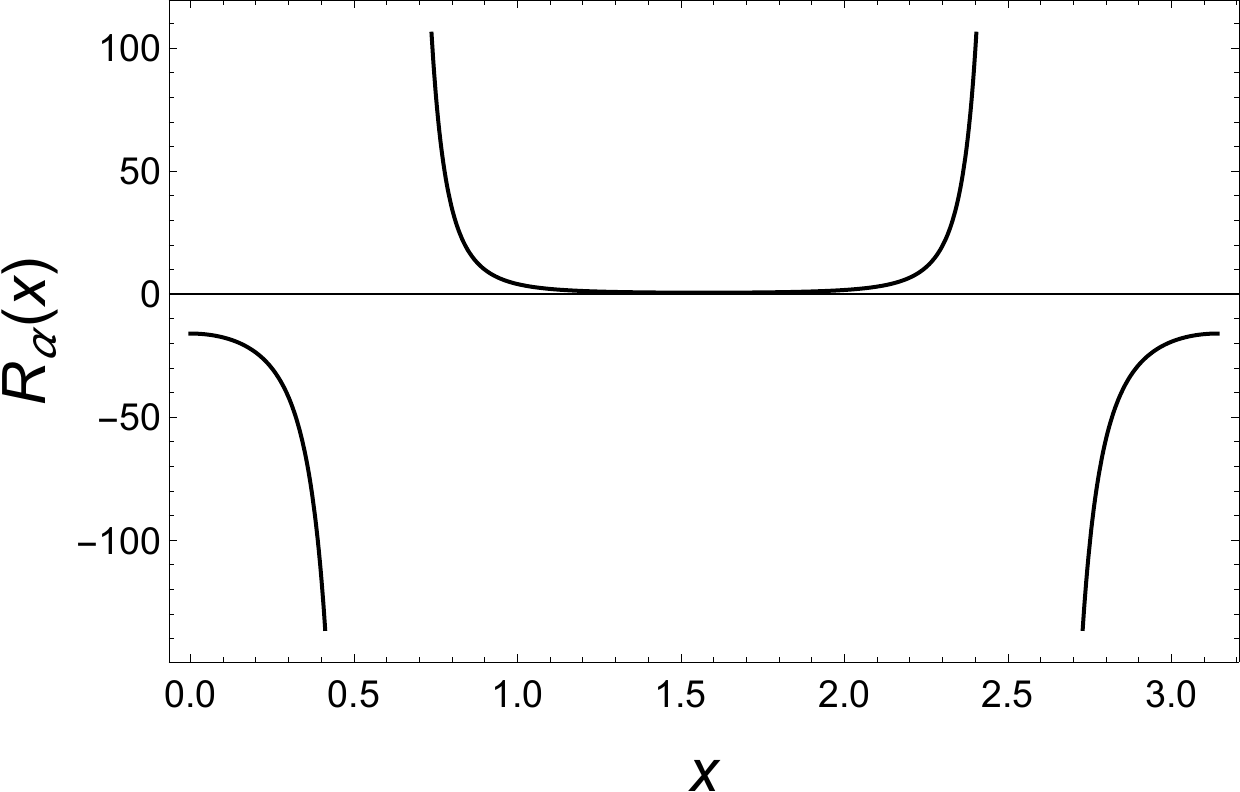}
\caption{The Rici curvature $R_\alpha$ as a function of the $x$ coordinate for $\alpha=\frac{3}{4}$. $R_\alpha$ becomes negative in the north and the  south regions of the sphere ($x\sim 0,\pi$). The singularities transition  takes place  at $x=\frac{1}{2} \arccos \left(\frac{7}{18}\right)$ and $x=\pi -\frac{1}{2} \arccos \left(\frac{7}{18}\right)$.}\label{191}
\end{center}
\end{figure}

The derivative with respect to $x^\mu=(x,y)$ of $\Lambda_h$ gives the tangent basis to the sphere at the point of coordinates $(x,y)$:
\begin{equation}\label{Ebasis}
E_\mu=\partial_\mu\Lambda_h=(-1)^{(\mu+1)}\textrm{A} f_{\mu'}\,e_1+\textrm{A}  f_{\mu"}\,e_2-\delta_{\mu,1}\textrm{B} \sin(x)\,e_3,
\end{equation}
with $\mu'=7-3\mu$ and $\mu"=4-\mu$. The basis $E_\mu$ permits us to construct the deformed metric $g_{\mu\nu}$ using the above scalar product $"\cdot "$. Fortunately the result is independent  of the variable $y$, namely
\begin{equation}\label{119}
g_{\mu\nu}=E_\mu\cdot E_\nu=\left[\sin^2(x)+(-\alpha^2)^{\delta_{2,\mu}}\cos^2(x)\right]\delta_{\mu\nu}-\alpha \epsilon_{\mu\nu}\cos(2x),
\end{equation}
where $\alpha=\tanh(h)$, $\delta_{\mu\nu}$ is the Kronecker symbol and $\epsilon_{\mu\nu}$ is the Levi-Civita anti-symmetric tensor. The metric inverse is given therefore by
\begin{equation}
g^{\mu\nu}=\left\{\left[\sin^2(x)+(-\alpha^2)^{\delta_{1,\mu}}\cos^2(x)\right]\delta^{\mu\nu}+\alpha \epsilon^{\mu\nu}\cos(2x)\right\}d_\alpha(x),
\end{equation}
where  $$d_\alpha(x)=det(g_{\mu\nu})^{-1}=(\sin^2(x)-\alpha^2\cos^2(x)+\alpha^2\cos^2(2x))^{-1}.$$ 
It is important to notice first that the metric tensor is no longer symmetric and second that $\Lambda_h\cdot E_1=E_1\cdot\Lambda_h=0$ as is the case in the non deformed geometry of the sphere. But however, $\Lambda_h\cdot E_2=-E_2\cdot\Lambda_h=-\alpha\sin(2x)$ means that the classical orthogonality is broken for $h\neq0$. The left dual basis $E^{\mu}$ of the cotangent bundle is defined by
\begin{equation}
E^\mu\equiv g^{\mu\nu}\star E_\nu=(-1)^{(\nu+1)} \textrm{A}\, g^{\mu\nu}\star f_{\nu'}\,e_1+\textrm{A}\, g^{\mu\nu}\star f_{\nu"}\,e_2-\textrm{B}\,g^{\mu,1}\sin(x)\,e_3.
\end{equation}
The right dual basis  $\tilde{E}^{\mu}$ is obtained from the left one by the general substitution  $h$ by $-h$ (or $\alpha\rightarrow -\alpha$) which have the effect to flip the factors of the  star product and the scalar product. Following reference \cite{CTZX08} we introduce the connection elements $\Gamma_{\mu\nu}^\sigma$ \footnote{These components represent the connection derivative $\nabla_\mu$ on the basis $E_\nu$: $\nabla_\mu E_\nu=\Gamma_{\mu\nu}^\sigma\star E_\sigma$} and $\Gamma_{\mu\nu\sigma}$ by
\begin{equation}
\Gamma_{\mu\nu}^\sigma=\partial_\mu E_{\nu}\cdot E^\sigma,\qquad \Gamma_{\mu\nu\sigma}=\partial_\mu E_{\nu}\cdot E_\sigma.
\end{equation}
The connection $\Gamma_{\mu\nu}^\sigma$ and the dual connection $\tilde{\Gamma}_{\mu\nu}^\sigma$  are related to the classical Christoffel  symbol $_c\Gamma_{\mu\nu\sigma}$ and the noncommutative torsion $\Upsilon_{\mu\nu\sigma}$ by
\begin{equation}
\Gamma_{\mu\nu}^\sigma=\left(_c\Gamma_{\mu\nu\alpha}+\Upsilon_{\mu\nu\alpha}\right)\star g^{\alpha\sigma}, \qquad\tilde{\Gamma}_{\mu\nu}^\sigma=g^{\sigma\alpha}\star \left(_c\Gamma_{\mu\nu\alpha}-\Upsilon_{\mu\nu\alpha}\right),
\end{equation}
where
\begin{equation}
_c\Gamma_{\mu\nu\sigma}=\frac{1}{2}(\partial_\mu g_{\nu\sigma}+\partial_\nu g_{\sigma\mu}-\partial_\sigma g_{\nu\mu}), \qquad \Upsilon_{\mu\nu\sigma}=\frac{1}{2}(\partial_\mu E_{\nu}\cdot E_\sigma-E_\sigma\cdot\partial_\mu E_{\nu}).
\end{equation}
The components of $\Gamma_{\mu\nu\sigma}$ are listed below
\begin{equation}
\begin{array}{l}
  \Gamma_{111}=0, \\ \Gamma_{112}= \Gamma_{222}=+\alpha\sin(2x), \\
  \Gamma_{121}= \Gamma_{211}=-\alpha\sin(2x),\\ \Gamma_{122}=\Gamma_{212}= \frac{1}{2}(1+\alpha^2)\sin(2x),\\
  \Gamma_{221}=-\frac{1}{2}(1+\alpha^2)\sin(2x).
\end{array}
\end{equation}
 Those of $\Gamma_{\mu\nu}^\sigma$ are
\begin{equation}\label{conex}
\begin{array}{l}
\Gamma_{\;11}^1=-\frac{\alpha^2}{2}\,d_\alpha(x)\sin(4x),\\
\Gamma_{\;11}^2=\alpha\, d_\alpha(x) \sin(2x),\\
\Gamma_{\;12}^1=\Gamma_{\;21}^1=\frac{1}{2}(\alpha^3-\alpha)d_\alpha(x)\sin(2x),\\
\Gamma_{\;12}^2=\Gamma_{\;21}^2=\frac{1}{2}(1+\alpha^2-2\alpha^2\cos(2x))d_\alpha(x)\sin(2x),\\
\Gamma_{\;22}^1=\frac{1}{4}(\cos(2x)(\alpha^2-1)^2+\alpha^4-1)d_\alpha(x)\sin(2x),\\
\Gamma_{\;22}^2=\alpha(1-\frac{1}{2}(1+\alpha^2)\cos(2x))d_\alpha(x)\sin(2x).
\end{array}
\end{equation}
Since these components are independent of $y$, we can easily compute the left Riemann tensor $R^l_{\;kij}$ using the formula
\begin{equation}
R^l_{\;kij}=-\partial_j\Gamma_{\;ik}^l-\Gamma_{\;ik}^p\star\Gamma_{\;jp}^l+\partial_i\Gamma_{\;jk}^l+\Gamma_{\;jk}^p\star\Gamma_{\;ip}^l.
\end{equation}
We are interest only by quoting the exact left star expressions of the Ricci curvature $R_{ij}$ and the scalar curvature $R$ which are defined respectively by
\begin{equation}
R_{ij}=R^p_{\;ipj},\qquad R=g^{ji}\star R_{ij}.
\end{equation}
So, we write
\begin{equation}\label{RicciTensor11}
R_{11}=\frac{3 \alpha ^4+4 \alpha ^2+1}{\left(\alpha ^2+2 \alpha ^2 \cos (2 x)-1\right)^2},
\end{equation}
\begin{equation}\label{RicciTensor12}
R_{12}= \frac{\alpha  \left(1-\alpha ^4\right) (\cos (2 x)+2)}{\left(\alpha ^2+2 \alpha ^2 \cos (2 x)-1\right)^2},
\end{equation}
\begin{equation}\label{RicciTensor21}
R_{21}= \frac{\alpha   \left(2 \left(\alpha ^2+1\right)^2+\left(\alpha ^4-1\right) \cos (2 x)\right)}{\left(\alpha ^2+2 \alpha ^2 \cos (2 x)-1\right)^2},
\end{equation}
\begin{equation}\label{RicciTensor22}
R_{22}=\frac{\left(1-\alpha ^4\right) \left(3 \alpha ^2+\left(3 \alpha ^2-1\right) \cos (2 x)+1\right)}{2 \left(\alpha ^2+2 \alpha ^2 \cos (2 x)-1\right)^2},
\end{equation}
and
\begin{equation}\label{RicciScalar}
R=\frac{2 \left(\alpha ^4-1\right) \left(3 \alpha ^2+2 \alpha ^2 \cos (2 x)+1\right)}{\left(\alpha ^2+2 \alpha ^2 \cos (2 x)-1\right)^3}.
\end{equation}
 The right star expressions are found as usual by the substitution $\alpha\rightarrow -\alpha$. Expressions (\ref{RicciTensor11})-(\ref{RicciTensor22}) and (\ref{RicciScalar}) generalize for arbitrary order of $\alpha$ or $h$ the results found in \cite{CTZX08} dealing with the case of noncommutative sphere. If we suppose $h\in \mathbb{R}$, the real parameter $\alpha=\tanh(h)$ ranges continuously from $-1$ to $+1$ and all above expressions are analytical in $\alpha$ for the almost values of $x$ such that ${\alpha ^2+2 \alpha ^2 \cos (2 x)-1}\neq 0$. This suggests the nice propriety that the right tangent module and the left tangent module are embedded in the same vector bundle equipped with the continuous basis $\{E_\mu(\alpha)\}$ over the space $S\times]-1,+1[$. In the extreme case $\alpha=\pm 1$ the curvature vanishes and the quantum sphere becomes flat everywhere on the map of the sphere. It is easy to show that the scalar curvature  could also  be negative for some specific regions on the sphere fig.\ref{191}.  When the quantum parameter is purely imaginary: $h=i \hbar$, the scalar curvature takes the form
\begin{equation}
\overline{R}=\frac{2 \left(\overline\alpha ^4-1\right) \left(3 \overline\alpha ^2+2 \overline\alpha ^2 \cos (2 x)-1\right)}{\left(\overline\alpha ^2+2 \overline\alpha ^2 \cos (2 x)+1\right)^3},
\end{equation}
where $\overline \alpha=\tan \hbar$. 
   Finally notice that at $\alpha=\overline{\alpha}=0$ the Ricci curvature reduces to the constant value $R=\overline R=2$ as expected for a commutative $2d$ sphere.
\section{Parallel transport field equation on the NC-sphere}\label{210}
According to different works based on different quantum deformation definitions, the notion of geodesic lines and the flow geodesic  on noncommutative space are not unique \cite{Parthaguha}, \cite{Borowiec}, \cite{Beggs}, \cite{Aschieri} , \cite{Golse}, \cite{KORDYUKOV} and \cite{Bolsinov}. An intuitive  approach is to consider the parallel transport equation $\nabla_XY=0$ for some two vector fields $X$ and $Y$ ($\nabla_X$ stands for the covariant derivative projected on the vector field $X$). When these two fields  are identified  to velocity the integral curve  is locally a geodesic path.  For the  purpose to define the noncommutative analog of the parallel transport equation we introduce the following expression based on Moyal product defined above
\begin{equation}\label{pGeod}
W^\mu \star D_\mu V^\sigma \star E_\sigma =W^\mu\star(\partial_\mu V^\sigma+V^\rho\star\Gamma_{\;\mu\rho}^\sigma)\star E_\sigma=0,
\end{equation}
where $V^\mu=V^\mu(x,y)$ and $W^\mu=W^\mu(x,y)$ are the field components on the tangent basis $E_\mu$ of the noncommutative sphere, $\Gamma_{\;\mu\rho}^\sigma$ are given by (\ref{conex}). The auto-parallel transport equation is a special case when $W^\mu=V^\mu$:
\begin{equation}\label{Geod}
V^\mu \star D_\mu V^\sigma \star E_\sigma =V^\mu\star(\partial_\mu V^\sigma+V^\rho\star\Gamma_{\;\mu\rho}^\sigma)\star E_\sigma=0.
\end{equation}
In fact the two equations (\ref{pGeod}) and (\ref{Geod})  are quite  different. The first one is a linear differential equation in $V^\mu$ and where $W^\mu$ is an arbitrary fixed field. The second is a non linear differential equation.  Before considering the treatment of the general case we are going to get insights into special cases of auto-parallel equation (\ref{Geod}).
\subsection{The $y$ independent fields: $y$-symmetry case}\label{204b}
Let suppose the vector field components  are real valued functions and independent of $y$ : $V^\mu(x,y)=V^\mu(x)=(\Psi(x),\Phi(x))\in \mathbb{R}$. After using the explicit components of the basis (\ref{Ebasis}), we obtain from (\ref{Geod}) a system of three equations.
\begin{equation}\label{204}
\displaystyle{\left\{
\begin{array}{l}
\Sigma^1_{\alpha}(x)\star f_4-\Sigma^2_{\alpha}(x)\star f_1=0,\\
\Sigma^1_{\alpha}(x)\star f_3+\Sigma^2_{\alpha}(x)\star f_2=0,\\
\Sigma^1_{\alpha}(x)\sin(x)=0,
\end{array}
\right.}
\end{equation}
where we have put $\Sigma^{\sigma}_{\alpha}(x)=V^\mu(\partial_\mu V^\sigma+V^\rho\Gamma_{\;\mu\rho}^\sigma)$ and $\Sigma^{\sigma}_{\pm}=\Sigma^{\sigma}_{\alpha}(x\pm i h)$ for in below purpose. Taking into count the formulas (\ref{Ffm}), the last system reduces to
\begin{equation}
\displaystyle{\left\{
\begin{array}{l}
\Sigma^1_+( f_4+i f_3)+\Sigma^1_-( f_4-i f_3)-\Sigma^2_+( f_1-i f_2)-\Sigma^2_-( f_1+i f_2)=0,\\
\Sigma^1_+( f_3-i f_4)+\Sigma^1_-( f_3+i f_4)+\Sigma^2_+( f_2+i f_1)+\Sigma^2_-( f_2-i f_1)=0,\\
\Sigma^1_{\alpha}(x)=0.
\end{array}
\right.}
\end{equation}
Multiplying the second equation by $i$ and adding and subtracting it from the first equation leads to the system
\begin{equation}\label{sysfirst}
\displaystyle{\left\{
\begin{array}{l}
\Sigma^1_+\cos(x)+i\Sigma^2_+\sin(x)=0,\\
\Sigma^1_-\cos(x)-i\Sigma^2_-\sin(x)=0,\\
\Sigma^1_{\alpha}(x)=0.
\end{array}
\right.}
\end{equation}
 where the substitutions $( f_1\pm i f_2)=\pm i \sin(x)e^{\mp i y}$, $( f_4\pm i f_3)= \cos(x)e^{\pm i y}$ and a global factorization of  $e^{\pm i y}$ are performed. Let us, notice that the third equation $\Sigma^1_{\alpha}(x)=0$ can be  obtained  by a limit  process ($h\rightarrow 0$) together with $\alpha$ fixed on the added two first ones\footnote{At this stage $\alpha$ and $h$ are treated independently even they are related by $\tanh$ function. Similarly we have  $\Sigma^2_{\alpha}(x)=0$ obtained by the same limit on the difference of the two first equations in (\ref{sysfirst}).}; and hence will be considered just as a constraint and will be omitted from the above system. Explicitly,  for any $( x,y)\in \left[0,\;\pi\right[\times\left[0,\;2\pi\right[$ and fixed $h$, and after using the identities $\cos(x)= \left(\cos(z)\pm i\alpha \sin(z)\right)/A$ and $\sin(x)= \left(\sin(z)\mp i\alpha \cos(z)\right)/A$ with $z=x\pm ih$ and $A$ given in (\ref{103}),  we have
 \begin{equation}\label{sysTwo}
\displaystyle{\left\{
\begin{array}{l}
\left.\left(\Sigma^1_{\alpha}(z)+\alpha\Sigma^2_{\alpha}(z)\right)\cos(z)+i\left(\Sigma^2_{\alpha} (z)+\alpha\Sigma^1_{\alpha}(z)\right)\sin(z)\;\right\vert_ {z=x+ih}=0,\\
\left.\left(\Sigma^1_{\alpha}(z)+\alpha\Sigma^2_{\alpha}(z)\right)\cos(z)-i\left(\Sigma^2_{\alpha} (z)+\alpha\Sigma^1_{\alpha}(z)\right)\sin(z)\;\right\vert_ {z=x-ih}=0.
\end{array}
\right.}
\end{equation}
 This form is an important non local differential equations system  for the components ($\Psi(x),\Phi(x)$). 
 In fact the  two equations in (\ref{sysTwo}) are distinct and are evaluated on different branes or  slices: $z=x\pm ih$ in $S\times]-1,1[$ showing the character of non local interaction between $V^\mu$ components resulting from the noncommutative effect on the  geometrical coordinates of the sphere. By the virtue of the connection (\ref{conex}), $\Sigma^1$ and $\Sigma^2$ have the following expressions
 \begin{equation}
\begin{array}{l}
\displaystyle{\Sigma^1_{\alpha}(x)= \Psi{\partial_1 \Psi}-\left\lbrace\alpha^2\cos(2x)\Psi^2-\alpha(\alpha^2-1)\Psi\Phi- \right.}
\\
\qquad\qquad\qquad\displaystyle{\left.\frac{1}{4}(\alpha^4-1+ (\alpha^2-1)^2\cos(2x))\Phi^2\right\rbrace d_\alpha(x)\sin(2x)},\\
\displaystyle{ \Sigma^2_{\alpha}(x)= \Psi\partial_1 \Phi+\left\lbrace\alpha \Psi^2+(1+\alpha^2-2\alpha^2\cos(2x))\Psi\Phi+\right.}\\
\qquad\qquad\qquad\displaystyle{\left.\alpha \left(1-\frac{1}{2}(\alpha^2+1)\cos(2x)\right)\Phi^2\right\rbrace d_\alpha(x)\sin(2x)}.
\end{array}
\end{equation}
 We have also from (\ref{sysfirst}) the compatibility equation
 \begin{equation}
 \Sigma^1_{\alpha}(x+i h)\Sigma^2_{\alpha}(x-i h)+\Sigma^2_{\alpha}(x+i h)\Sigma^1_{\alpha}(x-i h)=0.
 \end{equation}
\begin{figure}[h]
\begin{center}
{\bf a)} \includegraphics[scale=0.4]{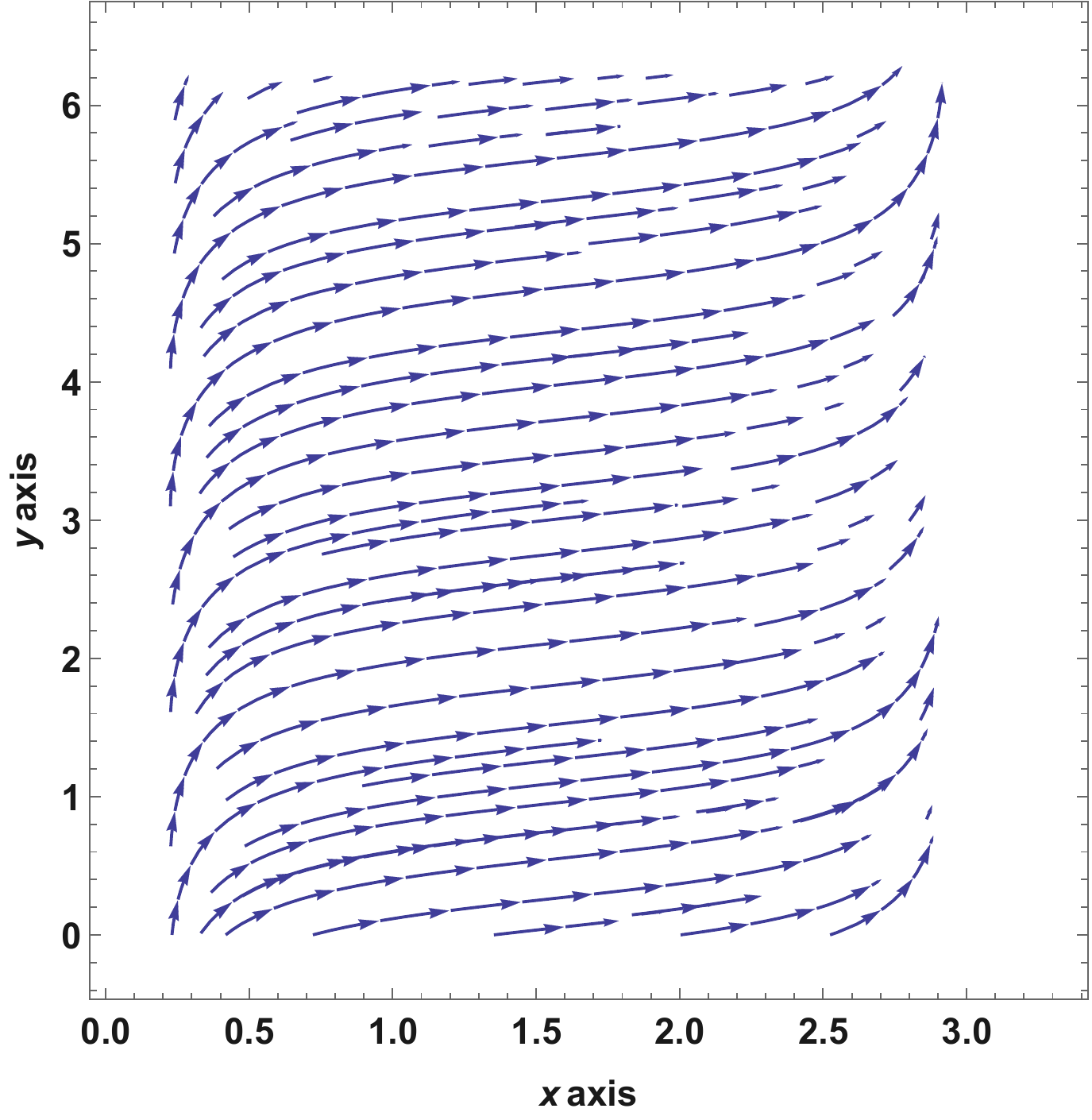} 
{\bf b)} \includegraphics[scale=0.35]{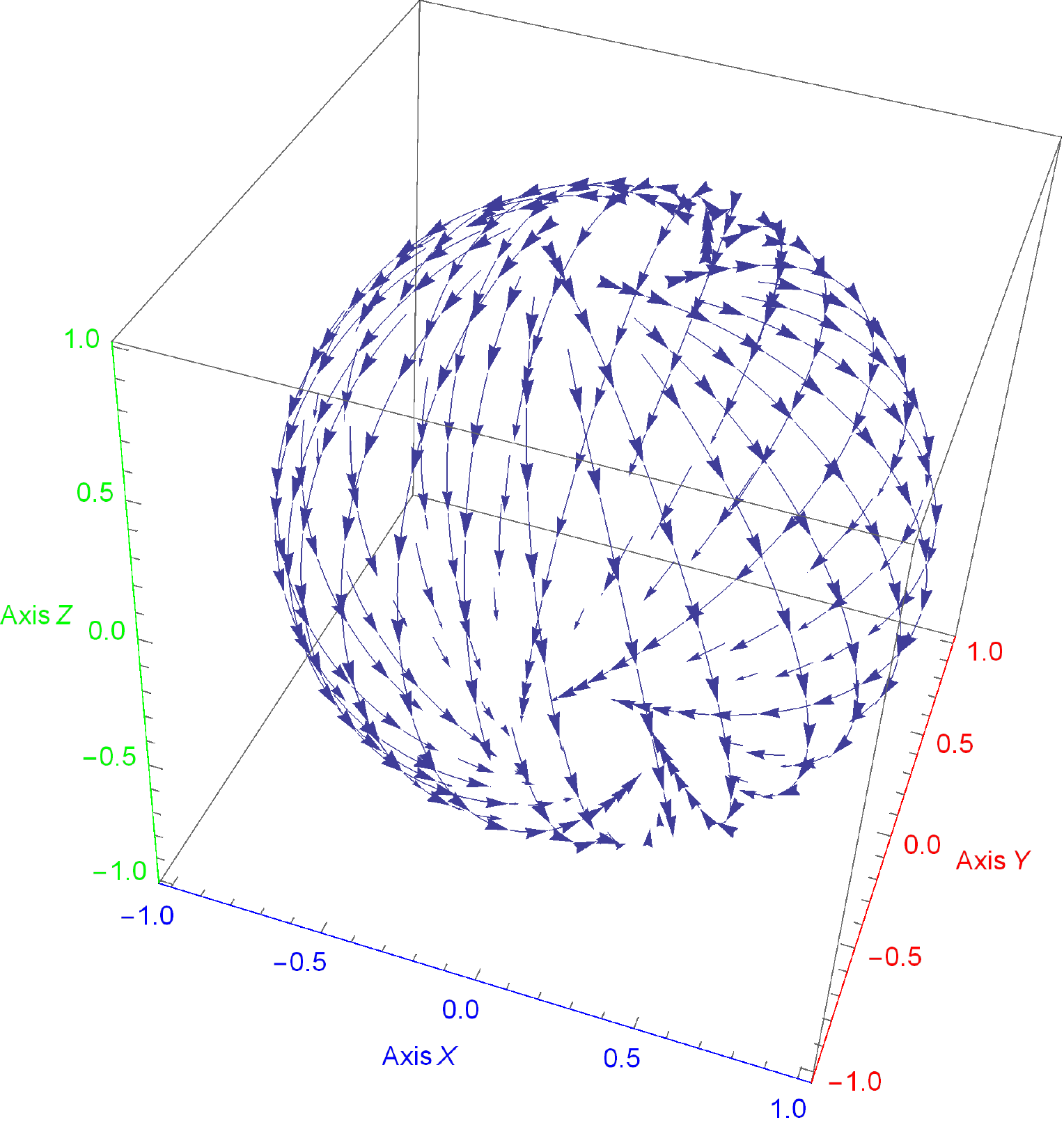}
\end{center}
\caption{\small {\bf a)} Plane representation of field flow (\ref{270}) on commutative sphere for a ratio $\frac{a_0}{b_0}=\frac{1}{20}$ and a spherical domain given by: $\arcsin \left( \sqrt{\frac{a_0}{b_0}}\right)\leq x \leq \pi-\arcsin \left( \sqrt{\frac{a_0}{b_0}}\right)$ and  $0\leq y \leq 2\pi$. {\bf b)} $3d$ representation on commutative sphere of the same field flow.}
\label{254}
\end{figure}
 The general solution of the system (\ref{sysTwo}) is not an easy task. For $\alpha =0$ our system is equivalent to a couple of  equations:  $\Sigma^1_0(x)=0$  and  $\Sigma^2_0(x)=0$ which explicitly read
 \begin{equation}\label{alphazerosys}
 \displaystyle{\left\{
\begin{array}{l}
\displaystyle{ \Psi\partial_1 \Psi-\frac{1}{2}\sin(2x)\Phi^2 =0},\\
\displaystyle{\Psi\partial_1 \Phi+2\Psi\Phi \cot (x)=0}.
\end{array}
\right.}
\end{equation}
The system (\ref{alphazerosys}) can be linearized by the substitution $\Phi_1=\Psi^2$, $\Phi_2=\Phi^2$ and the use of the variable  $\rho=\sin(x)^2$ \footnote{Let us remark that the system (\ref{alphazerosys}) is nonlinear (specially quadratic) and hence possesses the general discrete symmetry 
\begin{equation}\label{265}
\begin{array}{ll}
\Psi\longrightarrow \pm \Psi,& \Phi\longrightarrow \pm \Phi\\
\Psi\longrightarrow \pm i \Psi,& \Phi\longrightarrow \pm i \Phi.
\end{array}
\end{equation}
}. This leaves us with 
\begin{equation}\label{266}
\displaystyle{ \frac{d \Phi_1}{d \rho}=\Phi_2},\qquad\qquad\displaystyle{\rho\frac{d \Phi_2}{d \rho}=-2\Phi_2}.
\end{equation}
The solutions up to integration constants are $\Phi_1(\rho)=b_0-\frac{a_0}{\rho}=b_0-\frac{a_0}{\sin(x)^2}$ and $\Phi_2(\rho)=\frac{a_0}{\rho^2}=\frac{a_0}{\sin(x)^4}$. And finally the answer reads
\begin{equation}\label{270}
\psi_0(x)=\pm\sqrt{b_0-\frac{a_0}{\sin(x)^2}},\qquad \phi_0(x)=\pm\frac{\sqrt{a_0}}{\sin(x)^2}.
\end{equation}
 The reality of the field $( \psi_0(x),\phi_0(x))$ imposes the conditions: $a_0\geq 0$ and $1\geq \sin (x)^2\geq \frac{a_0}{b_0}$. In figure \ref{254} we have given the plot of this case when the above conditions are observed. In this sample the flow ($\sqrt{b_0-\frac{a_0}{\sin(x)^2}},\frac{\sqrt{a_0}}{\sin(x)^2}$)  starts as a  vortex in the northern region, crosses the latitude lines, then the equator and reaches the southern region. For $a_0\neq 0$ there are two forbidden zones or holes around south and north poles. Every line field belongs to the half greatest circle on the sphere as  expected from geodesic  proprieties. In the contrast case when conditions are not verified the field becomes imaginary function   and there is no real flow solution on the $2d$ sphere.

For the first order in $\alpha$ the system (\ref{sysfirst}) gives the following equations
\begin{equation}\label{279}
\left\lbrace
\begin{array}{r}
\Psi\partial_1\Psi-\frac{1}{2}\sin(2x) \Phi^2-\alpha \left\lbrace 2 \partial_1 (\Psi\Phi)+\tan(x)\left(\partial_1(\Psi\partial_1 \Phi)-2 \Phi \Psi\right)\right\rbrace  =0, \\
\\
 \tan(x) \Psi \partial_1 \Phi \!+\!2 \Psi \Phi \!+\! \alpha \left\lbrace 2 \Psi^2 \!+\!4 \sin(x)^2\Phi^2\!-\!\sin(2x)\Phi\partial_1 \Phi\!+\!(\partial_1 \Psi)^2\!+\! \Psi\partial_1^2 \Psi\right\rbrace = 0.
\end{array}\right.
\end{equation}
These are second order differential equations possessing two independent solutions in contrast to equations of the first system given in (\ref{alphazerosys}) which is reducible to first order differential system (see (\ref{266})) and possessing one solution. When  the parameter $\alpha$ is switched to  zero at least one solution of (\ref{279}) is confluent to the solution of (\ref{270}) modulo the discrete symmetry (\ref{265}). Let us call this the connected  solution and let  find its deviation from  (\ref{270}) for small values of $\alpha$. For this purpose let us insert
\begin{eqnarray}
&&\Psi=\psi_0+\alpha\; \Omega (x),\qquad\qquad \Phi=\phi_0+\alpha\; \Delta (x).
\end{eqnarray}
 in the last equations (\ref{279}) and keep only the first order in $\alpha$; After some simplification,  the final expression of the  system is then
 \begin{equation}\label{300}
 \left\lbrace
\begin{array}{l}
\left(\psi_0\;\partial_1 +\psi _0'\right)\Omega (x) -\sin (2 x)\phi _0\;\Delta (x) + \psi _0 \phi _0'= 0,\\
\\
\psi _0 \left( \tan(x) \; \partial_1 +2\right)  \Delta (x)+\phi _0^2+2 {b_0}= 0.
\end{array}
\right.
 \end{equation}
 This is a non homogeneous system  and is integrated by the constant variation method which gives the expressions  of  the  field deviation:
 \begin{equation}\label{300b}
 \left\lbrace
\begin{array}{l}
\Omega (x)\!=\!\displaystyle{d_0\frac{ \sin (x)}{\sqrt{\eta (x)}}\!+\!\sqrt{a_0}\left(c_0\frac{ 2\psi _0}{\eta (x)}\!+\!1\! -\! \frac{\sqrt{2\beta (x)}}{\sin (x)}\left(1\!-\!\frac{{a_0}}{\eta (x)}\right) \tan ^{-1}\left(\frac{\sqrt{2} \sin (x)}{\sqrt{\beta(x)}}\right)\right)},\!
\\
\Delta (x)\!=\!\displaystyle{\frac{1}{\sin(x)^2} \left(c_0-\frac{{a_0}}{\sqrt{{b_0}}} \coth ^{-1}\left(\frac{\psi _0}{\sqrt{b_0}}\right)-\left(1+\sin ^2(x)\right)\psi _0\right)},
\end{array}
\right.
 \end{equation}
where 
\begin{equation}
\eta (x)=2(a_0-b_0\sin (x)^2),\qquad \beta (x)=\frac{\eta (x)}{b_0}=-\frac{2}{b_0}\sin (x)^2 \psi_0^2
\end{equation} 
 and $c_0$ and $d_0$ are new integration constants. At this level we have a special situation to highlight. In fact for real positive parameters $a_0$ and $b_0$($>a_0$), the two squares $\sqrt{\eta (x)}=\sqrt{2(a_0-b_0\sin (x)^2)} $ and $\psi_0= \sqrt{b_0-\frac{a_0}{\sin(x)^2}}$ cannot  be simultaneously real which implies that   $\Psi (x) $ is a complex  valued function and henceforth cannot represent a physical flow unless a specific choice is made for the constants $c_0,d_0$ and $\alpha$. So in the generic case it is not possible to draw a realistic flow deviation  similar to figure \ref{254}. Of course the solution (\ref{300b}), which is interpreted as complex quantum fluctuations, still has the $y$-rotation symmetry around north-south axis of the sphere as well as the mirror inversion symmetry between north and south since the full expression depends   on $\sin (x)^2=\sin (\pi-x)^2$.
\subsection{General case of auto-parallel transport equation}\label{354}
As the vector fields are defined on spherical surface they must fulfill the symmetry requirements which are the special orthogonal rotation symmetries. It follows from the angular momentum symmetry  that the natural basis for smooth functions on a sphere surfaces is the one spanned by the spherical harmonics $Y_l^m(x,y)$. We can then assume that $V^\mu(x,y)$ for $x\in\left[0,\;\pi\right.\left[\right.$ and $y\in\left[0,\;2\pi\right.\left[\right.$ is square integral real function and  therefore can be uniquely decomposed on  spherical harmonics basis $Y_l^m(x,y)$:
\begin{equation}\label{VtoY}
V^\mu(x,y)=\sum_{l=0}^\infty\sum_{m=-l}^{+l}\left(C^\mu_{l,m}Y_l^m(x,y)+{C^{\mu\;*}_{l,m}}Y_l^{m\,*}(x,y)\right),
\end{equation}
where $C^\mu_{l,m}$ are some complex constants carrying the vector index $\mu$. Using the relation between spherical harmonics and Legendre polynomials $P_{l,m}(\cos(x))$, namely
\begin{equation}\label{336}
Y_l^m(x,y)=k_{l,m}P_{l,m}(\cos (x))\exp(i\,m\,y),
\end{equation}
where $k_{l,m}$ are explicitly  given by ${\sqrt{\frac{(2l+1)}{4\pi}\frac{(l-|m|)!}{(l+|m|)!}}}$
and $C^\mu_{l,m}=\frac{1}{2}(a^\mu_{l,m}+i\,b^\mu_{l,m})$ ($a$ and $b\in \mathbb{R}$), we re-express (\ref{VtoY}) as follow
\begin{equation}\label{347}
\begin{array}{l}
 \displaystyle{V^\mu(x,y)=\sum_{l=0}^\infty\sum_{m=-l}^{+l}k_{l,m}P_{l,m}(x)\left\{a^\mu_{l,m}\cos(m\,y)- b^\mu_{l,m}\sin(m\,y)\right\}}.
\end{array}
\end{equation}
We notice that variables $x$ and $y$ are separated  in  (\ref{336}) and hopefully  the $\exp(i\,m\,y)$ is an adequate function for the star product. Also the expression of (\ref{347}) shows after terms rearrangement a Fourier series behavior  if uniform convergence is assumed:
  \begin{equation}\label{351}
\begin{array}{l}
 \displaystyle{V^\mu(x,y)=V_0^\mu(x)+\sum_{n=1}^\infty \left\{V^\mu_{C,n}(x)\cos(n\,y)+V^\mu_{S,n}(x)\sin(n\,y)\right\}}.
\end{array}
\end{equation}
Here for instance the zero mode term reads  $\displaystyle{V_0^\mu(x)=\sum_{l=0}^\infty k_{l,0}P_{l,0}(x)a^\mu_{l,0}}$. To compute the Fourier coefficients we will use later  the scalar product
$\langle gh \rangle =\int_0^{2\pi}dy\,g(y)h(y)$ defined on the circle $S^1$. Now let insert (\ref{351}) in   (\ref{Geod}) to get the following system of equations 
\begin{equation}\label{363}
\displaystyle{\left\{
\begin{array}{l}
\hat\Sigma^1_{\alpha}(x,y)\star f_4-\hat\Sigma^2_{\alpha}(x,y)\star f_1=0,\\
\hat\Sigma^1_{\alpha}(x,y)\star f_3+\hat\Sigma^2_{\alpha}(x,y)\star f_2=0,\\
\hat\Sigma^1_{\alpha}(x,y)\star\sin(x)=0,
\end{array}
\right.}
\end{equation}
where this time we have put $\hat\Sigma^{\sigma}_{\alpha}(x,y)=V^\mu(x,y)\star(\partial_\mu V^\sigma(x,y)+V^\rho(x,y)\star\Gamma_{\;\mu\rho}^\sigma(x))$. The later expression according to (\ref{351}) is given by
\begin{eqnarray}
&& \hat\Sigma^{\sigma}_{\alpha}(x,y)=\left\lbrace V_0^\mu(x)+\sum_{n=1}^\infty \left\{c_n V^\mu_{C,n}(x)+s_n V^\mu_{S,n}(x)\right\}\right\rbrace \star\nb\\
&&\qquad\qquad\quad \left\lbrace {\cal C}_{0,\mu}^\sigma +\sum_{m=1}^{\infty}\left[c_m{\cal C}_{m,\mu}^\sigma (x,h) +s_m{\cal S}_{m,\mu}^\sigma (x,h)\right] \right\rbrace , \label{375}
\end{eqnarray}
where $c_n=\cos (n y)$ and $s_n=\sin (n y)$ and the Fourier mode functions are introduced:
\begin{eqnarray}
&& {\cal C}_{0,\mu}^\sigma \,=\! \delta_\mu ^1\partial_1 V_0^{\sigma}(x) + V_0 ^{\rho}(x) \Gamma _{\mu \rho}^{\sigma}(x),\nb \\
&&{\cal C}_{m,\mu}^\sigma \!=\!\displaystyle{\delta_\mu ^1\partial_1 V^\sigma_{C,m}\!+\!m \delta_\mu ^2 V^\sigma_{S,m}\!+\!\frac{1}{2}\left(\left(V^\rho_{C,m}\!+\!i\,V^\rho_{S,m}\right)\Gamma _{\mu \rho}^{\sigma}(x\!+\!i \,h)\!+\!\left(V^\rho_{C,m}\!-\!i\,V^\rho_{S,m}\right)\Gamma _{\mu \rho}^{\sigma}(x\!-\!i \,h)\right)},\nb\\
&&{\cal S}_{m,\mu}^\sigma \!=\!\displaystyle{\delta_\mu ^1\partial_1 V^\sigma_{S,m}\!-\!m \delta_\mu ^2 V^\sigma_{C,m}\!+\!\frac{1}{2}\left(\left(V^\rho_{S,m}\!-\!i\,V^\rho_{C,m}\right)\Gamma _{\mu \rho}^{\sigma}(x\!+\!i \,h)\!+\!\left(V^\rho_{S,m}\!+\!i\,V^\rho_{C,m}\right)\Gamma _{\mu \rho}^{\sigma}(x\!-\!i \,h)\right)}.\nb
\end{eqnarray}
 Now let expand the star product in  (\ref{375}). We obtain
\begin{eqnarray}
&& \hat\Sigma^{\sigma}_{\alpha}(x,y)=\Sigma^{\sigma}_{\alpha}(x)+\sum_{m=1}^{\infty}\left[(V_0^\mu\star c_m){\cal C}_{m,\mu}^\sigma  +(V_0^\mu\star s_m){\cal S}_{m,\mu}^\sigma +\right.\nb\\
&& \qquad\quad\quad
\left.
(c_m\star {\cal C}_{0,\mu}^\sigma )V^\mu_{C,m}+(s_m \star {\cal C}_{0,\mu}^\sigma )V^\mu_{S,m}
\right]+\sum_{n,m=1}^\infty \left[(c_n V^\mu_{C,n})\star (c_m {\cal C}_{m,\mu}^\sigma )+\right. \nb 
\\
&&\qquad\quad\quad\left.(c_n V^\mu_{C,n})\star (s_m {\cal S}_{m,\mu}^\sigma )+(s_n V^\mu_{S,n})\star (c_m {\cal C}_{m,\mu}^\sigma )+(s_n V^\mu_{S,n})\star (s_m {\cal S}_{m,\mu}^\sigma )\right]. \nb
\end{eqnarray}
The first term $\Sigma^{\sigma}_{\alpha}(x)$ is already defined just after equation (\ref{204}). Using  formulas (\ref{400})-(\ref{403}) and  (\ref{406})-(\ref{413}) of appendix~\ref{398} and \ref{398b} we return back to commutative  multiplication:
\begin{eqnarray}
&& \hat\Sigma^{\sigma}_{\alpha}(x,y)=\Sigma^{\sigma}_{\alpha}(x)+\frac{1}{2}\sum_{m=1}^{\infty}\left\lbrace c_m{\cal B}_{1,m}^{\;\;\;\sigma}(x)-s_m {\cal B}_{2,m}^{\;\;\;\sigma}(x)\right\rbrace +\sum_{n,m=1}^\infty\left\lbrace\right.
\nb\\
&&
\qquad \left. c_n c_m {\cal B}_{3,m}^{n,\sigma}(x)+c_n s_m {\cal B}_{4,m}^{n,\sigma}(x)+
 s_n c_m {\cal B}_{5,m}^{n,\sigma}(x)+s_n s_m {\cal B}_{6,m}^{n,\sigma}(x)\right\rbrace ,
\end{eqnarray}
where the ${\cal B}_j\,^\sigma(x)$ are defined by (\ref{450b})-(\ref{460}) in the appendix~\ref{444}.
 To achieve our computation we have to evaluate for $p\in {\mathbb{N}}\!-\!\{0\}$ the Fourier integrals 
\begin{equation}\label{410}
\int_0^{2\pi}dy \;c_p \;(\hat\Sigma^{\sigma}_{\alpha}(x,y)\star f_j),\qquad \int_0^{2\pi}dy \;s_p\;(\hat\Sigma^{\sigma}_{\alpha}(x,y)\star f_j),
\end{equation} 
   for $j=1,\cdots , 4$ and
\begin{equation}\label{412}
\int_0^{2\pi}dy\;c_p\;(\hat\Sigma^{\sigma}_{\alpha}(x,y)\star \sin (x)),\qquad \int_0^{2\pi}dy\;s_p\;(\hat\Sigma^{\sigma}_{\alpha}(x,y)\star \sin (x)),
\end{equation}  
to pick the $p$-modes from the system of equations (\ref{363}). The details are summarized in the appendix~\ref{454}. Finally, the $y$-integration of (\ref{363}) over cosine and sine modes will give, after algebraic simplifications, the following system

\begin{eqnarray}
&& \cos (x) \left[-({\cal K}_{1} - {\cal L}_{1})_{p-1}^{(1)} \texttt{ch}_{p-1}-({\cal K}_{1} - {\cal L}_{1})_{p-1}^{(2)} \texttt{sh}_{p-1}+4\delta _{1,p} (\Sigma (w)+\Sigma (z))_{\alpha }^1\right]+\nb\\
&& \quad\sin (x) \left[-({\cal K}_{2} + {\cal L}_{2})_{p-1}^{(2)} \texttt{ch}_{p-1}-({\cal K}_{2} + {\cal L}_{2})_{p-1}^{(1)} \texttt{sh}_{p-1}+
4i\delta _{1,p} (\Sigma (w)-\Sigma (z))_{\alpha }^2\right]=0,\nb\\
&& \cos (x) \left[-({\cal K}_{2} + {\cal L}_{2})_{p-1}^{(1)} \texttt{ch}_{p-1}-({\cal K}_{2} + {\cal L}_{2})_{p-1}^{(2)} \texttt{sh}_{p-1}+4i\delta _{1,p} (\Sigma (w)-\Sigma (z))_{\alpha }^1\right]+\nb\\
&&\quad \sin (x) \left[+({\cal K}_{1} - {\cal L}_{1})_{p-1}^{(2)} \texttt{ch}_{p-1}+({\cal K}_{1} - {\cal L}_{1})_{p-1}^{(1)} \texttt{sh}_{p-1}-4\delta _{1,p} (\Sigma (w)+\Sigma (z))_{\alpha }^2\right]=0,\nb
\\
&&\cos (x) \left[{\cal L}_{4,p+1}^{1} \texttt{ch}_{p+1}-{\cal L}_{4,p+1}^{2} \texttt{sh}_{p+1}\right]-\sin (x) \left[i{\cal L}_{3,p+1}^{2} \texttt{ch}_{p+1}-i{\cal L}_{3,p+1}^{1} \texttt{sh}_{p+1}\right]=0\nb\\
&&\cos (x) \left[i{\cal L}_{3,p+1}^{1} \texttt{ch}_{p+1}-i{\cal L}_{3,p+1}^{2} \texttt{sh}_{p+1}\right]+\sin (x) \left[{\cal L}_{4,p+1}^{2} \texttt{ch}_{p+1}-{\cal L}_{4,p+1}^{1} \texttt{sh}_{p+1}\right]=0,\nb
\\
&&\hspace{2cm} {\rm ch}_p\sin (x)({\cal B}_{1}+{\cal M}_{1})_{p}^{1}\!+\!{\rm sh}_p\cos (x)({\cal B}_{2}-{\cal M}_{2})_{p}^{1}=0,\nb\\
&&\hspace{2cm}{\rm sh}_p\cos (x)({\cal B}_{1}+{\cal M}_{1})_{p}^{1}\!-\!{\rm ch}_p \sin (x)({\cal B}_{2}-{\cal M}_{2})_{p}^{1}=0,\label{422}
\end{eqnarray}
where the functions ${\cal K}_{j}(x)$, ${\cal L}_{j}(x)$ and ${\cal M}_{j}(x)$ are defined in subsections \ref{454}.

The last system (\ref{422}) of differential equations  represents our final result. It describes the $p$ Fourier mode ($V^\mu_{0}(x),V^\mu_{C,p}(x),V^\mu_{S,p}(x)$) of the auto parallel field on the quantum sphere. We can deduce from the last two equations of (\ref{422}) the simple constraints 
\begin{equation}
{\cal M}_{1,p}^{1}=-{\cal B}_{1,p}^{1},\quad
{\cal M}_{2,p}^{1}={\cal B}_{2,p}^{1}
\end{equation}
which show a deep coupling  between different modes of $x$ component of the field. For $p=1$ the two first equations of (\ref{422}) reduce to the $y$ independent system (\ref{sysfirst}). The great challenge is how to solve this system of non-local equations. Such objective is beyond the scope of the actual work.
\section{Results and comments} \label{448}
In this work we were interested on $2d$ quantum gravity  introduced on spherical surface via the  noncommutative Moyal star product. We have established equations describing quantum effect on the geodesic flow equation. For the first goal we succeed to obtain the exact quantum expressions for the compenents of the Levi Cevita connection, the Ricci tensor and the scalar curvature, generalizing  those of \cite{CTZX08}. We found that  the quantum sphere becomes flat ($R=0$) in the extreme case  when $h$ goes to $\infty$ ($\alpha=\pm 1$). Moreover, the scalar curvature could even, for some values of $\alpha$, be negative on some region of the sphere, a fact  which could have an impact on the geodesic lines  if solutions exist there. 

For the second goal we have resolved the auto parallel field equation at the zero order and the first order of the quantum parameter $\alpha$. Under the assumption  of $y$-symmetry the   solution for the first order is a complex field almost everywhere. This means  that this correction couldn't be interpreted as physical geodesic  deviation  at least for this order of approximation. Our general result (\ref{422}) for a given mode $p=1,2,...\infty$  shows  an interdependence between  higher and lower modes. Moreover the finite gap between $w=x+ih$ and $z=x-ih$ (for finite value of $h$) makes these equations non local differential equations system. The $y$-symmetry situation studied in subsection \ref{204b} is the case where all higher "oscillations" are desactivated. The strategy to solve (\ref{422}) is to activate the few $N$ first modes  and switch-off the remaining infinite modes (${\cal B}_j=0$ when $n,m>N$) then proceed to $\alpha$ expansion of the differential equations. Evidently the $y$-symmetry will break and technical difficulties need more care. At any order of $\alpha$, spherical rotations around arbitrary axis of the sphere offer  a way to generate new solutions.

We have to notice that our actual results are not intrinsic proprieties of the sphere but depends on the choice of the local chart. In fact we have  used the spherical system coordinate to describe points of sphere: $x=\theta$ for latitude and $y=\phi$ for longitude. Also we have choose a star product based on Heinsenberg quantization rule ($[x,y]=h$) and obviously this product is  not covariant under local diffeomorphisms of the sphere. In the expression (\ref{103b}) defining the vector $\Lambda_h$ we have made a symmetric choice that gave us the  metric  (\ref{119}) which is independent of $y$ and it is this fact that made the subsequent computations more easy. Other asymmetric choice of $\Lambda_h$ are  more difficult even if they fulfill the classical condition of the unit radial vector $\Lambda_0=f_2\,e_1+\,f_1\,e_2+\cos(x)\,e_3$ when $h$ goes to $0$. The last remark to make concerns the use both of the usual product $"\cdot "$ and  the star product $"\star "$ inside the Fourier integrals (\ref{410}) and (\ref{412}). A possible alternative is to use only the star product but this way will ask for more efforts and will give an alternative flow equations system.

This work could be reproduced for quantum $2d$-hyperbolic surfaces just by changing the signature of scalar product on ${\cal A}^3$ (see section \ref{97}) as follows $[a,b,c]\cdot[a',b',c']=a\star a'+b\star b'-c\star c'$ and using the Wick rotation: $x\rightarrow i\, x$ in the trigonometric functions. The results are formally very similar to the spherical case  and don't need more investigations. However physical implications could  be different and interesting as the chaotic regime of the flow on hyperbolic surface. Other important questions remain without response in our work. They  concern first the effect of the quantum deformation on geodesic flow beyond the first order of $\alpha$ and second what is about the other solutions not confluent to~(\ref{270}). As soon perspective the quantum torus $T^2$ will be treated to show how geodesic flow will behave under quantum deformations. The $S^3$ sphere viewed as hypersurface in euclidean four dimension space is another  good candidate to apply the formalism of noncommutative surfaces. The two last examples are so important because both are Lie groups and parallelizable manifolds and we expect that the geodesic flow is unaltered by "suitable canonical quantization" of their  map  coordinates. Other schemes of quantization are possible for the star product; for instance we cite the star product based on Lie commutators ($[x,y]=i h z,...$). But computation are not easy to conduct for these schemes and important efforts should be developed.   

\subsubsection*{Acknowledgment}
I would like to express my thanks to Dr. Noureddine Bouayed at the physics department of Blida1 University for useful discussion and remarks, and to Prof. Dominique Manchon of  "{Laboratoire de Math\'ematiques  Blaise Pascal.}" for very appreciated reflexions  and observations by emails and during my presence in Blaise Pascal University. Also I would like to thank my friend Brahim Mohammed ZAHAF, from "D\'epartement de math\'ematiques, Laboratoire d'Analyse Non Lin\'eaire et Math\'ematiques Appliqu\'ees (LANLMA), Universit\'e de  Tlemcen", for reading and making comments on this work.

\appendix 
\section{Some useful star products and mode integration}\label{398a} 
\subsection{Star product of $F(x)$ with trigonometric functions}\label{398} 
Let us quote some relations for star product including trigonometric functions
 \begin{eqnarray}
 && 2 F(x)\star c_m = (c_m-i s_m)F_- + (c_m+i s_m)F_+,\label{400}\\
 && 2 F(x)\star s_m = (s_m-i c_m)F_+ +(s_m+i c_m)F_-, \\
 && 2 c_m\star F(x)= (c_m-i s_m)F_+ +(c_m+i s_m)F_-, \\
 && 2 s_m\star F(x)=(s_m-i c_m)F_- +(s_m+i c_m)F_+ ,\label{403}
\end{eqnarray}
where $F_\pm =F(x\pm i h m)$ for short notation. 
\subsection{Reduced $\star$-product of $F(x)$ and $G(x)$}\label{398b} 
The star product which concerns the non commutative variables $x$ and $y$ induces a star product on functions of $x$ only  when trigonometric functions are implicated as follow  
 \begin{eqnarray}
 (c_n F)\!\star\! (c_m G)\!=\! A_{FG,1}^{(nm)} c_n c_m\!+\! A_{FG,2}^{(nm)} c_n s_m \!+\!A_{FG,3}^{(nm)} s_n c_m \!+\!A_{FG,4}^{(nm)}  s_n s_m ,\label{406}\\
 (c_n F)\!\star\! (s_m G)\!=\! A_{FG,1}^{(nm)}  c_n s_m\!-\! A_{FG,2}^{(nm)} c_n c_m  \!+\!A_{FG,3}^{(nm)} s_n s_m \!-\!A_{FG,4}^{(nm)} s_n c_m,\\
 (s_n F)\!\star\! (c_m G)\!=\! A_{FG,1}^{(nm)}  s_n c_m\!+\! A_{FG,2}^{(nm)}  s_n s_m\!-\!A_{FG,3}^{(nm)}  c_n c_m\!-\!A_{FG,4}^{(nm)}  c_n s_m,\\
 (s_n F)\!\star\! (s_m G)\!=\! A_{FG,1}^{(nm)}  s_n s_m \!-\! A_{FG,2}^{(nm)} s_n c_m \!-\!A_{FG,3}^{(nm)}c_n s_m \!+\!A_{FG,4}^{(nm)}  c_n c_m,\label{413}
 \end{eqnarray}
where we have introduced $4$ new residual star products of $F$ and $G$ on $x$ dimension only:
 \begin{eqnarray}
 &&A_{FG,1}^{(nm)}=+\cos(n h \partial_{\otimes 1})\cos(m h\partial_{\otimes 2})(F\otimes G)(x),\label{443}\\
 &&A_{FG,2}^{(nm)}=-\cos(n h \partial_{\otimes 1})\sin(m h\partial_{\otimes 2})(F\otimes G)(x),\\
 &&A_{FG,3}^{(nm)}=+\sin(n h \partial_{\otimes 1})\cos(m h\partial_{\otimes 2})(F\otimes G)(x),\\
 &&A_{FG,4}^{(nm)}=-\sin(n h \partial_{\otimes 1})\sin(m h\partial_{\otimes 2})(F\otimes G)(x),\label{450}
 \end{eqnarray}
where $\partial_{\otimes 1}=\partial_1\otimes 1$ (resp. $\partial_{\otimes 2}=1\otimes\partial_1$) acts as a derivative on $F(x)$ (resp. $G(x)$)of the tensor product.   We call these expressions  {\it reduced $\star$-products}.
\subsection{Definitions of ${\cal B}_j\,^\sigma(x)$ functions}\label{444}
For short notation we have used above the following expressions
\begin{eqnarray}
&& {\cal B}_{1,m}^{\;\;\;\sigma}(x)\!=\! 
( V_{0-}^{\mu}\!+\! V_{0+}^\mu) {\cal C}_{m,\mu}^\sigma  \!-\!i(V_{0+}^\mu-
V_{0-}^\mu){\cal S}_{m,\mu}^\sigma  \!+\!
\nb\\
&&\qquad\qquad({\cal C}_{0+,\mu}^\sigma \!+\! {\cal C}_{0-,\mu}^\sigma) V^\mu_{C,m}\!-\!i ({\cal C}_{0-,\mu}^\sigma \!-\!{\cal C}_{0+,\mu}^\sigma) V^\mu_{S,m}\label{450b},\\
&& 
{\cal B}_{2,m}^{\;\;\;\sigma}(x)\!=\!i(V_{0-}^{\mu}\!-\!V_{0+}^\mu){\cal C}_{m,\mu}^\sigma \! -\!(V_{0+}^\mu +V_{0-}^\mu){\cal S}_{m,\mu}^\sigma \!+\!i({\cal C}_{0+,\mu}^\sigma \!-\!
\nb\\
&&\qquad\qquad {\cal C}_{0-,\mu}^\sigma)V^\mu_{C,m}  \!-\!({\cal C}_{0-,\mu}^\sigma \!+\! {\cal C}_{0+,\mu}^\sigma )V^\mu_{S,m},\\
 && {\cal B}_{3,m}^{n,\sigma}(x)=A_{CC,1}^{(nm)}-A_{CS,2}^{(nm)}-A_{SC,3}^{(nm)}+A_{SS,4}^{(nm)},\\
 && {\cal B}_{4,m}^{n,\sigma}(x)=A_{CC,2}^{(nm)}+A_{CS,1}^{(nm)}-A_{SC,4}^{(nm)}-A_{SS,3}^{(nm)},\\
 && {\cal B}_{5,m}^{n,\sigma}(x)=A_{CC,3}^{(nm)}-A_{CS,4}^{(nm)}+A_{SC,1}^{(nm)}-A_{SS,2}^{(nm)},\\
 && {\cal B}_{6,m}^{n,\sigma}(x)=A_{CC,4}^{(nm)}+A_{CS,3}^{(nm)}+A_{SC,2}^{(nm)}+A_{SS,1}^{(nm)},  \label{460} 
 \end{eqnarray}
where the subscripts $\{CC,CS,SC,SS\}$ of $A_{FG,i}$ mean that the substitution 
$$\{(V^\mu_{C,n},{\cal C}_{m,\mu}^\sigma),(V^\mu_{C,n},{\cal S}_{m,\mu}^\sigma),(V^\mu_{S,n},{\cal C}_{m,\mu}^\sigma),(V^\mu_{S,n},{\cal S}_{m,\mu}^\sigma)\}$$
 is understood in the reduced $\star$-products (\ref{443})-(\ref{450}).  
\subsection{$p$-Modes of the star products $\hat\Sigma^{\sigma}_{\alpha}\star \sin(x)$ and $\hat\Sigma^{\sigma}_{\alpha}\star f_j. $}\label{454}

With $p>0$ and  the abbreviated notation ${\rm sh}_p=\sinh (p h)$ and ${\rm ch}_p=\cosh ( p h)$ we have 
\begin{eqnarray}
\int_0^{2\pi}dy\, c_p(\hat\Sigma^{\sigma}_{\alpha}(x,y)\star \sin (x))\!=\!
\frac{\pi}{2}\left\lbrace {\rm ch}_p\sin (x)({\cal B}_{1}+{\cal M}_{1})_{p}^{\sigma}\!+\!{\rm sh}_p\cos (x)({\cal B}_{2}-{\cal M}_{2})_{p}^{\sigma}\right\rbrace ,\nb
&& \\ \int_0^{2\pi}dy\,s_p(\hat\Sigma^{\sigma}_{\alpha}(x,y)\star \sin (x))=
\frac{\pi}{2}\left\lbrace{\rm sh}_p\cos (x)({\cal B}_{1}+{\cal M}_{1})_{p}^{\sigma}\!-\!{\rm ch}_p \sin (x)({\cal B}_{2}-{\cal M}_{2})_{p}^{\sigma}\right\rbrace ,\nb &&
\end{eqnarray}
where 
\begin{eqnarray}
&&{\cal M}_{1,p}^{\sigma}(x)=\sum_{n=1}^\infty ({\cal B}_{3,{p-n}}^{n,\sigma} \!-\! {\cal B}_{6,{p-n}}^{n,\sigma}\!+\!{\cal B}_{3,{p+n}}^{n,\sigma} \!+\!{\cal B}_{6,{p+n}}^{n,\sigma}\!+\!{\cal B}_{3,{n-p}}^{n,\sigma} \!+\!{\cal B}_{6,{n-p}}^{n,\sigma}),\nb\\
&&{\cal M}_{2,p}^{\sigma}(x)=\sum_{n=1}^\infty ( {\cal B}_{4,{p+n}}^{n,\sigma}\!-\!{\cal B}_{5,{p+n}}^{n,\sigma}\!+\!{\cal B}_{4,{p-n}}^{n,\sigma}\!+\!{\cal B}_{5,{p-n}}^{n,\sigma}\!-\!{\cal B}_{4,{n-p}}^{n,\sigma}\!+\!{\cal B}_{5,{n-p}}^{n,\sigma}).\nb
\end{eqnarray}
Next, setting  $z=x+i h$, $w=x-ih$ and knowing that ${\cal B}_{j,m=0}^{n,\sigma}=0$ we have 
 \begin{eqnarray}
&&\int_0^{2\pi} c_p \hat\Sigma^{\sigma}_{\alpha}(x,y)\!\star\! f_1\!=\!\frac{\pi}{8}   \left\lbrace
\texttt{ch}_{p-1} ({\cal K}_2 \!+\!{\cal L}_2)^{\sigma}_{p-1}\sin (x)\!+\!
   \texttt{sh}_{p-1}({\cal K}_1\!-\!{\cal L}_1)^{\sigma}_{p-1} \cos (x)+\right.
\nb \\
&&\left.
i \texttt{ch}_{p+1} {\cal L}^{\sigma}_{3,p+1}\sin(x)\!+\!\texttt{sh}_{p+1}{\cal L}^{\sigma}_{4,p+1} \cos (x)\right\rbrace \!+\!\frac{i \pi}{2}  
   \sin (x) \delta _{1,p} \left(\Sigma _{\alpha }^{\sigma }(z)\!-\!\Sigma _{\alpha }^{\sigma }(w)\right)
 ,\\
&&\int_0^{2\pi} c_p\hat\Sigma^{\sigma}_{\alpha}(x,y)\!\star\! f_2\!=\!\frac{\pi}{8} \left\lbrace\texttt{sh}_{p-1} ({\cal K}_2 \!+\!{\cal L}_2)^{\sigma}_{p-1}\cos(x)\!-\!\texttt{ch}_{p-1} ({\cal K}_1-{\cal L}_1)^{\sigma}_{p-1} \sin (x)\!+\!\right.
\nb\\
&&\left.\texttt{ch}_{p+1} {\cal L}^{\sigma}_{4,p+1} \sin (x)-i \texttt{sh}_{p+1} {\cal L}^{\sigma}_{3,p+1} \cos   (x)\right\rbrace+\frac{\pi}{2}  \sin (x) \delta _{1,p} \left(\Sigma _{\alpha }^{\sigma }(w)+\Sigma _{\alpha }^{\sigma
   }(z)\right)
,\\
&& \int_0^{2\pi} c_p\hat\Sigma^{\sigma}_{\alpha}(x,y)\star f_3=\frac{\pi}{8}   \left\lbrace\texttt{ch}_{p-1} ({\cal K}_2 +{\cal L}_2)^{\sigma}_{p-1}\cos (x)-\texttt{sh}_{p-1}({\cal K}_1-{\cal L}_1)^{\sigma}_{p-1}  \sin (x)+\right.
\nb\\
&&\left. i \texttt{ch}_{p+1} {\cal L}^{\sigma}_{3,p+1} \cos(x)- \texttt{sh}_{p+1}{\cal L}^{\sigma}_{4,p+1}  \sin (x)\right\rbrace+\frac{\pi }{2} i 
   \cos (x) \delta _{1,p} \left(\Sigma _{\alpha }^{\sigma }(z)-\Sigma _{\alpha }^{\sigma }(w)\right),
\end{eqnarray}
 \begin{eqnarray}
&& \int_0^{2\pi} c_p\hat\Sigma^{\sigma}_{\alpha}(x,y)\star f_4=\frac{\pi}{8}\left\lbrace -\texttt{ch}_{p-1} ({\cal K}_1-{\cal L}_1)^{\sigma}_{p-1} \cos(x)-\texttt{sh}_{p-1}({\cal K}_2 +{\cal L}_2)^{\sigma}_{p-1} \sin (x)+\right.
\nb\\
&&\left.\texttt{ch}_{p+1}{\cal L}^{\sigma}_{4,p+1} \cos (x)+i \texttt{sh}_{p+1}{\cal L}^{\sigma}_{3,p+1} \sin(x)\right\rbrace +\frac{\pi}{2}\cos (x) \delta _{1,p} \left(\Sigma _{\alpha }^{\sigma }(w)+\Sigma _{\alpha }^{\sigma
   }(z)\right)
,\\
&&\int_0^{2\pi} s_p \hat\Sigma^{\sigma}_{\alpha}(x,y)\star f_1=\frac{\pi}{8}\left\lbrace \texttt{sh}_{p-1}({\cal K}_2 +{\cal L}_2)^{\sigma}_{p-1} \cos(x)-\texttt{ch}_{p-1} ({\cal K}_1-{\cal L}_1)^{\sigma}_{p-1}\sin (x)-\right.
\nb\\
&&\left.
\texttt{ch}_{p+1}{\cal L}^{\sigma}_{4,p+1} \sin (x)+i  \texttt{sh}_{p+1}{\cal L}^{\sigma}_{3,p+1} \cos(x)\right\rbrace+\frac{\pi }{2}  \sin (x) \delta _{1,p} \left(\Sigma _{\alpha }^{\sigma }(w)+\Sigma _{\alpha }^{\sigma}(z)\right)
,\\
&&\int_0^{2\pi} s_p\hat\Sigma^{\sigma}_{\alpha}(x,y)\star f_2=\frac{\pi}{8}\left\lbrace -\texttt{ch}_{p-1}({\cal K}_2 +{\cal L}_2)^{\sigma}_{p-1} \sin (x)-\texttt{sh}_{p-1} ({\cal K}_1-{\cal L}_1)^{\sigma}_{p-1} \cos (x)+ \right.
\nb\\
&&\left.i \texttt{ch}_{p+1} {\cal L}^{\sigma}_{3,p+1}\sin(x)+\texttt{sh}_{p+1} {\cal L}^{\sigma}_{4,p+1}\cos (x)\right\rbrace-\frac{\pi}{2}i\sin (x) \delta _{1,p} \left(\Sigma _{\alpha }^{\sigma }(z)-\Sigma _{\alpha }^{\sigma }(w)\right)
,\\
&& \int_0^{2\pi} s_p\hat\Sigma^{\sigma}_{\alpha}(x,y)\star f_3=\frac{\pi}{8}\left\lbrace -\texttt{ch}_{p-1}({\cal K}_1-{\cal L}_1)^{\sigma}_{p-1}\cos(x)-\texttt{sh}_{p-1}({\cal K}_2 +{\cal L}_2)^{\sigma}_{p-1}\sin (x)-\right.
\nb\\
&&\left.\texttt{ch}_{p+1}{\cal L}^{\sigma}_{4,p+1} \cos (x)-i \texttt{sh}_{p+1} {\cal L}^{\sigma}_{3,p+1}\sin(x)\right\rbrace+\frac{\pi}{2}\cos (x) \delta _{1,p} \left(\Sigma _{\alpha }^{\sigma }(w)+\Sigma _{\alpha }^{\sigma}(z)\right)
,\\
&& \int_0^{2\pi} s_p\hat\Sigma^{\sigma}_{\alpha}(x,y)\star f_4=\frac{\pi}{8}\left\lbrace \texttt{sh}_{p-1}({\cal K}_1-{\cal L}_1)^{\sigma}_{p-1} \sin (x)-\texttt{ch}_{p-1}({\cal K}_2 +{\cal L}_2)^{\sigma}_{p-1} \cos (x)+\right.
\nb\\
&&\left. i \texttt{ch}_{p+1}{\cal L}^{\sigma}_{3,p+1}\cos(x)- \texttt{sh}_{p+1} {\cal L}^{\sigma}_{4,p+1}\sin (x)\right\rbrace-\frac{\pi}{2} i\cos (x) \delta _{1,p} \left(\Sigma _{\alpha }^{\sigma }(z)-\Sigma _{\alpha }^{\sigma }(w)\right),
\end{eqnarray}
where 
\begin{eqnarray}
&&{\cal K}^{\sigma}_{1,p}=\sum_{n=1}^{\infty}\left[-{\cal B}_{3}(w)-i {\cal B}_{4}(w)+i {\cal B}_{5}(w)-{\cal B}_{6}(w)-{\cal B}_{3}(z)+i{\cal B}_{4}(z)-i {\cal B}_{5}(z)-{\cal B}_{6}(z)\right]_{|n-p|}^{n,\sigma }
\nb\\
&&\qquad\quad\quad\;\;\left[-{\cal B}_{3}(w)+i {\cal B}_{4}(w)-i {\cal B}_{5}(w)-{\cal B}_{6}(w)-{\cal B}_{3}(z)-i {\cal B}_{4}(z)+i {\cal B}_{5}(z)-{\cal B}_{6}(z)\right]_{n+p}^{n,\sigma }
,\nb\\
&&{\cal K}^{\sigma}_{2,p}=\sum_{n=1}^{\infty}\left[-i {\cal B}_{3}(w)+{\cal B}_{4}(w)-{\cal B}_{5}(w)-i {\cal B}_{6}(w)+i {\cal B}_{3}(z)+{\cal B}_{4}(z)-{\cal B}_{5}(z)+i{\cal B}_{6}(z)\right]_{|n-p|}^{n,\sigma }
\nb\\
&&\qquad\quad\quad\;\;\left[-i {\cal B}_{3}(w)-{\cal B}_{4}(w)+{\cal B}_{5}(w)-i{\cal B}_{6}(w)+i {\cal B}_{3}(z)-{\cal B}_{4}(z)+{\cal B}_{5}(z)+i {\cal B}_{6}(z)\right]_{n+p}^{n,\sigma }
,\nb\\
&&\qquad\qquad {\cal L}^{\sigma}_{1,p}=\left[+{\cal B}_{1 }(w)+i {\cal B}_{2}(w)+{\cal B}_{1}(z)-i {\cal B}_{2 }(z)\right]_{p}^{\sigma }
,\nb\\
&&\qquad\qquad{\cal L}^{\sigma}_{2,p}=\left[-i {\cal B}_{1}(w)+{\cal B}_{2}(w)+i {\cal B}_{1}(z)+{\cal B}_{2}(z)\right]_{p}^{\sigma }
,\nb\\
&&\qquad\qquad {\cal L}^{\sigma}_{3,p}=\left[-{\cal B}_{1}(w)+i {\cal B}_{2}(w)+{\cal B}_{1}(z)+i {\cal B}_{2}(z)\right]_{p}^{\sigma }
,\nb\\
&&\qquad\qquad{\cal L}^{\sigma}_{4,p}=\left[+{\cal B}_{1}(w)-i {\cal B}_{2}(w)+{\cal B}_{1}(z)+i {\cal B}_{2}(z)\right]_{p}^{\sigma }.\nb
\end{eqnarray}


\begin{thebibliography}{200}
\bibitem{CTZX08} M. Chaichian, A. Tureanu, R.B. Zhang, X. Zhang, {\it Riemannian geometry of noncommutative surfaces }, J. Math. Phys. {\bf 49}, 073511 (2008), {\tt arXiv:hep-th/0612128 }
\bibitem{SW} N. Seiberg, E. Witten, J. High Energy Phys. {\bf 09}, 32 (1999), {\tt arXiv:hep-th/9908142 }
\bibitem{SD} S. Doplicher, Proceedings of the 37th Karpacz Winter School of Theoretical Physics, 2001, 204-213, {\tt arXiv:hep-th/0105251v2 }.
\bibitem{SDF} S. Doplicher, K. Fredenhagen, J.E. Roberts, Commun. Math.
Phys. {\bf 172}, 187 (1995), {\tt arXiv:hep-th/0303037 }.
\bibitem{KLV} Kanazawa, T., Lambiase, G., Vilasi, G. et al. Eur. Phys. J. C (2019) {\bf 79}: 95, https://doi.org/10.1140/epjc/s10052-019-6610-1
\bibitem{Koba} A. Kobakhidze, Int. J. Mod. Phys. {\bf A 23}, 2541 (2008), {\tt arXiv:hep-th/0603132 }.
\bibitem{Cham} A.H. Chamseddine, Phys. Lett. B {\bf 504}, 33 (2001), {\tt arXiv:hep-th/0009153 }.
\bibitem{Djabari} L. Bonora, M. Schnabl, M. Sheikh-Jabbari, A. Tomasiello, Nucl. Phys. B {\bf 589}, 461 (2000),  {\tt arXiv:hep-th/0006091 }.
\bibitem{chaichian} M. Chaichian, P. Presnajder, A. Tureanu, Phys. Rev. Lett. {\bf 94}, 151602 (2005), {\tt arXiv:hep-th/0409096 }.
\bibitem{wess} P. Aschieri, C. Blohmann, M. Dimitrijevic, F. Meyer, P. Schupp, J. Wess, Class. Quantum Gravity {\bf 22}, 3511 (2005),  {\tt arXiv:hep-th/0504183 }.
\bibitem{Modesto} L. Modesto and P. Nicolini, Phys. Rev. D {\bf 81}, 104040 (2010), {\tt arXiv:hep-th/0912.0220 }.
\bibitem{Nicolini}P. Nicolini, Noncommutative Black Holes, The Final Appeal To Quantum Gravity: A Review, Int. J. Mod. Phys. {\bf A 24}, 1229 (2009), {\tt arXiv:hep-th/0807.1939 }. 
\bibitem{Moffat} J. W. Moffat, Noncommutative quantum gravity, Phys. Lett. {\bf B 491}, 345 (2000), {\tt arXiv:hep-th/0007181 }.
\bibitem{Calmet2} X. Calmet and A. Kobakhidze, Noncommutative general relativity, Phys. Rev. {\bf D 72}, 045010 (2005), {\tt arXiv:hep-th/0506157 }. 
\bibitem{Calmet3} X. Calmet and A. Kobakhidze, Second order noncommutative corrections to gravity, Phys. Rev. {\bf D 74}, 047702 (2006), {\tt arXiv:hep-th/0605275 }.
\bibitem{Garay} L. J. Garay, (1994). Quantum Gravity and Minimum Length. Int. J. Mod. Phys. {\bf A10} (1995) 145-166, arXiv: {\tt gr-qc/9403008 }.
\bibitem{Hinchliffe} I. Hinchliffe, N. Kersting and Y. L. Ma, Review of the Phenomenology of Noncommutative Geometry. Int. J. Mod. Phys. {\bf A. 19.} No. 02 pp. 179-204 (2004), {\tt arXiv:hep-ph/0205040 }.
\bibitem{Freidel} L. Freidel and E. R. Livine, Phys. Rev. Lett. {\bf 96}, (2006) 221301 , 3D Quantum Gravity and Effective Noncommutative Quantum Field Theory, {\tt arXiv:hep-th/0512113 }.
\bibitem{Garcia} H. Garc\'{i}a-Compe\'{a}n, O. Obreg\'{o}n and C.Ram\'{i}rez, Noncommutative Quantum Cosmology, Phys.Rev.Lett.{\bf 88}, 161301 (2002),  {\tt arXiv:hep-th/0107250 }.
\bibitem{chaichian2} M. Chaichian, A. Tureanu, G. Zet, Phys. Lett. B {\bf 660}, 573 (2008), {\tt arXiv:hep-th/0710.2075 }. 
\bibitem{Calmet} X. Calmet and G. Landsberg, Lower Dimensional Quantum Black Holes, Published in chapter 7 in A.J. Bauer and D.G.Eiffel editors,Black Holes: Evolution, Theory and Thermodynamics Nova Publishers, New York, 2012,  {\tt arXiv:hep-th/1008.3390 }.
\bibitem{wheeler} R. Ruffini and J. A. Wheeler, Introducing the black hole, Phys. Today 24, 30 (1971),  https://doi.org/10.1063/1.3022513
\bibitem{Renteln} Paul Renteln,  Manifolds, Tensors, and Forms: An Introduction for Mathematicians and Physicists. Cambridge Univ. Press. (2013), p. 328. ISBN 1107659698.
\bibitem{Parthaguha} Partha Guha,. (2011). Geodesic flows, Von Neumann  equation  and  quantum  mechanics on noncommutative cylinder. Mod. Phys. Let. {\bf A. 21.}, https://doi.org/10.1142/S0217732306020305
\bibitem{Borowiec} A. Borowiec, Vector fields and differential operators: noncommutative case, Czech.
J. Phys. {\bf 47} (1997) 1093-1100, {\tt arXiv:q-alg/9710006 }.
\bibitem{Beggs} E. Beggs, Noncommutative geodesics and the KSGNS construction, {\tt arXiv:math-qa/1811.07601v2  }.
\bibitem{Aschieri} P. Aschieri and P. Schupp, Vector fields  on Quantum Groups, Int. J. Mod. Phys. {\bf A
11} (1996) 1077-1100, https://doi.org/10.1142/S0217751X9600050X
\bibitem{Golse} F. Golse and E. Leichtman, Applications of Connes' Geodesic Flow
to Trace Formulas in Noncommutative Geometry, Journal of Functional Analysis {\bf 160}, 408-436 (1998, https://doi.org/10.1006/jfan.1998.3305
\bibitem{KORDYUKOV} Y. A. Kordyukov, Egorov’s Theorem for Transversally Elliptic Operators on Foliated Manifolds and Noncommutative Geodesic Flow, Math. Phys. Anal. and Geom. (2005) {\bf 8}: 97–119,  {\tt arXiv:math/0407435v1 }. 
\bibitem{Bolsinov} A. V. Bolsinov,  and  B. Jovanović, Non-commutative integrability, moment map and geodesic flows, Ann. Glob. Anal. Geom. (2003) {\bf 23}: 305, {\tt arXiv:math-ph/0109031 }. 
\end{thebibliography}
\end{document}